\documentclass[conference]{IEEEtran}
\IEEEoverridecommandlockouts
\usepackage{cite}
\usepackage{amsmath,amssymb,amsfonts}
\usepackage{algorithmic}
\usepackage{graphicx}
\usepackage{textcomp}
\usepackage{xcolor}
\usepackage[T1]{fontenc}
\def\BibTeX{{\rm B\kern-.05em{\sc i\kern-.025em b}\kern-.08em
    T\kern-.1667em\lower.7ex\hbox{E}\kern-.125emX}}
    
\begin{document}

\title{The Influence of Social Networks on Human Society\\
}

\author{\IEEEauthorblockN{ Shreyash Arya}
\IEEEauthorblockA{shreyash15097@iiitd.ac.in}
}

\maketitle

\begin{abstract}
This report gives a brief overview of the origin of social networks and their most popular manifestation in the modern era - the Online Social Networks (OSNs) or social media. It further discusses the positive and negative implications of OSNs on the human society. The coupling of Data Science and social media (social media mining) is then put forward as a powerful tool to overcome the current challenges and pave the path for futuristic advancements.

\end{abstract}


\section{Introduction}
The origin and application of graph theory in 1726 for solving the Köningsberg Problem \cite{b1}, along with a combination of sociology, statistics, and social psychology, introduced the modeling of any social system as a graph. A social network is a social graph formed by the connection between entities (individuals or organizations) and their interactions. In today’s modern world, social networks have been adopted in multiple domains ranging from communication, demographics, economics, linguistics, advertisement, complex contagion, diffusion, evolution, and the most common and popular adaptation - the Online Social Networks (OSNs) or Social Media. This report highlights the current standings of these OSNs along with its influence and future in modern human society.


The introduction of the internet on mobile devices has truly revolutionized the way we communicate and share information and social media being the catalyst, approximately half of the world’s population is virtually connected on these social media platforms \cite{b2}. People have become more connected than ever before, and OSNs, being the carriers, have become an essential commodity. According to the “six degrees of separation” theory, any person is a maximum of six degrees socially distant if one forms a chain using the “friend of a friend” rule \cite{b3}. The gap between any two people can be thus bridged in just a few taps or clicks, and with instant messaging applications such as WhatsApp and Facebook messenger, this notion of separation also fades away.   

In this report, I first provide an overview of the positive and negative influences of OSNs on human society, in Section II. Subsequently, Section III discusses the role of Data Science in harnessing the power of OSNs. Finally, Section IV concludes the report and gives an outlook on future work.

\section{Influence of Online Social Networks}

\subsection{The Positives}
Through \textbf{better connectivity}, social media has helped in improving cross-cultural communication, building social and technical skills, and providing social support to improve health literacy among adolescents and cognitive abilities among older adults \cite{b8,b9,b10,b11,b12}. \textbf{Politically}, people have become more involved in highlighting important issues (such as the Tunisian revolution 2011), amplifying their political views by organizing protests, spreading the word online, and tracking social media footprints of candidates during elections \cite{b23, b24, b25,b26}. Not only individuals, but \textbf{organizations and governments} are also using it for better interaction with the citizens, promoting open government platforms, and analyzing the opinions of their citizens \cite{b13}. 

Social media is based on the principles of Web 2.0, where user-generated content and user interactions are its lifeblood \cite{b4,b5}. It offers a plethora of opportunities to \textbf{content creators} to create different content-type, be it in the form of photos (Instagram, Pinterest), videos (YouTube, TikTok), microblogging (Twitter), or even game streaming (Twitch) \cite{b27,b28,b29}. Further, the primary \textbf{source of information} - news, has become a participation work through sharing, and about 62\% of adults in the United States use social media (Facebook and Twitter) as a daily news source \cite{b6,b7}. For \textbf{businesses}, social media platforms provide a way to reach customers in the form of targeted advertisements, learning their users’ behaviors and interests for personalized recommendations of items, and boosting promotion reach using influential entities on the network \cite{b14,b15,b16,b17}. 

The above-discussed examples are just the tip of the iceberg. With numerous possible applications and incorporation of advanced technology such as Artificial Intelligence (AI) into social media, there are a lot of potential applications that will help in the betterment of society. But there are indeed two sides to the story.

\subsection{The Negatives}
Social media is criticized for its adverse impacts on society. There has been significant \textbf{information disparity} attacking net neutrality (treating all internet communications equal) by providing an unfair advantage to companies like Facebook to increase their coverage/usage over others \cite{b30,b31}.

\textbf{Privacy and security}: The data we share in the form of photos, video, text, or location on social media creates a snapshot of our lives. Even with the privacy policies and settings in place, data can be accessed without consent by third-party applications, government agencies, or by using social media mining and intelligence techniques\cite{b51,b53,b54,b55}. People are ignorant of their actions online as it has become a norm to share their personal lives and contribute to the social network even for the sake of privacy. There have been cases where teens share their emails, phone numbers, addresses, and other compromising information, inviting online predators, harassment, trolling, catfishing, stalking, stereotyping, cyberbullying, and internet fraud \cite{b40,b41,b42,b43,b44,b45,b46,b47,b48,b49,b50,b51,b52,b56,b57,b58}.

\textbf{Health} (physical, mental, and emotional) is another major concern negatively related to social media use. The urge of virtual gratification in the form of likes, comments, and followers, comparison of online social life, and peer pressure are the key reasons leading to disorders related to body perception and eating habits along with depression, anxiety, disturbed sleep, mood swings, less satisfaction with life, and social media addiction  \cite{b60,b61,b62,b63,b64,b65,b66,b67,b68,b69,b70,b71}. 


In \textbf{political} scenario, there has been polarization with selective exposure to information due to created echo-chambers (confined regions of singular amplified thoughts), persuasion of political beliefs and misinformation/pseudo information/fake news \cite{b32,b33,b34,b35,b36,b37,b38}. Also, the ease of accessing social media has promoted slacktivism - supporting a political or social cause with little commitment and effort, which in turn undermines the true essence of such movements \cite{b39}.

It seems very evident that with all the negatives overpowering the positives, it is the need of the hour to rectify the flaws as social media has immense positive potential for the present and future, and this is where Data Science (discussed in the next section) comes to the rescue.

\section{A Ray of Hope: Data Science - Social Media Mining and Analysis}

Data Science, an interdisciplinary field used to extract knowledge and insights from data, when clubbed with social media is called social media mining. It is put to use in many domains: in \textbf{health care} - detecting cancer using Twitter, networked smoking cessation, and drug-drug interaction and reactions; in \textbf{education} - evaluation of content such as academic publications, Wikipedia articles, and resumes; in \textbf{privacy and security} - detection of misinformation, collusion, fake profiles, sybils, bots, and cyborgs, location-based privacy, provenance of information, political polarization; in \textbf{business and marketing} - targeted advertisement, personalized recommendation of products, services and users; and in \textbf{disaster/crisis management} - detecting earthquakes and its surrounding rumors \cite{b75,b76,b77,b78,b79,b80,b81,b82,b83,b84,b85,b86,b87,b88,b89,b90,b91,b92,b37,b38,b93,b94,b95,b96,b97,b98,b99}. 

In a nutshell, Data Science combined with social media is a powerful tool capable of 1) providing solutions to the current challenges discussed in section II.B, 2) learning the intricacies and  patterns  of  social  media  to  improve  and  innovate  over the current versions and 3) enabling the intersection of newest technological breakthroughs such as AI with social media.

\section{Conclusion and Future Scope}

This report provides a brief overview of the origin of social networks, its current incarnation in the form of OSNs or social media, and its implications on human society. It seems very evident with the examples discussed in section II.A and II.B that the benefits of better communication, information awareness, and freedom to create and share content, all can be overweighed by the drawbacks of improper usage, handling, and loopholes of the system. There are many solutions proposed, but the most effective is using Data Science combined with social media (called social media mining) to overcome the challenges and pave a way to the advancement of social media. Thus, \textit{“With great power comes great responsibility”}, the famous aphorism by comic writer Stan Lee also applies to the use of social networks and social media. Social media is ubiquitous and has immense potential for the future. We will be noticing futuristic technological changes on these platforms using Artificial Intelligence, Augmented Reality, and Virtual Reality, in addition to the rise of social influencers, social commerce, and enhanced privacy \cite{b100,b101,b102,b103,b104,b105,b106}. In the end, rather than debating the positives or negatives of social media, we need to look for domains or use cases it is good for and for what it is not. Coupled with data science, OSNs can be better analyzed and controlled to ensure a safe and functional environment for all users.


\begin{thebibliography}{00}
\bibitem{b68} "How social media reflects our daily mood changes". Medical News Today.
\end{thebibliography}
\end{document}